\documentclass[12pt]{article}
\newcommand{\beq}{\begin{equation}}
\newcommand{\eeq}{\end{equation}}
\newcommand{\beqa}{\begin{eqnarray}}
\newcommand{\eeqa}{\end{eqnarray}}
\newcommand{\beqar}{\begin{eqnarray*}}
\newcommand{\eeqar}{\end{eqnarray*}}

\newcommand{\inn}{\!\cdot\!}

\newcommand{\la}{\lambda}
\newcommand{\Lam}{\Lambda}

\newcommand{\ie}{{\it i.e.,}\ }
\newcommand{\labell}[1]{\label{#1}} %{\label{#1}} %
\newcommand{\reef}[1]{(\ref{#1})}
\newcommand\prt{\partial}

\newcommand\hD{\hat{D}}

\newcommand\tD{{\tilde D}}
\newcommand\tR{{\tilde R}}
\newcommand\tP{{\tilde P}}

\newcommand\tQ{{\tilde Q}}

\newcommand\Tr{{\rm Tr}}
\newcommand\STr{{\rm STr}}
\begin{document}

 \vspace*{1cm}

\begin{center}
{\bf \Large
On non-linear action of multiple M2-branes

 }
\vspace*{1cm}

{Mohammad R. Garousi}\\
\vspace*{0.2cm}
{ Department of Physics, Ferdowsi university, P.O. Box 1436, Mashhad, Iran}\\
\vspace*{0.1cm}
%and\\
{ School of Physics, 
 Institute for research in fundamental sciences (IPM), \\
 P.O. Box 19395-5531, Tehran, Iran. 
}\\
\vspace*{0.4cm}

\vspace{2cm}
ABSTRACT
\end{center}
A nonlinear  $SO(8)$ invariant BF type Lagrangian for describing the dynamics  of N   M2-branes in flat spacetime has been proposed recently in the literature which is an extension of the non-abelian DBI action of N D2-branes. This action includes only terms with even number of the totally antisymmetric tensor $M^{IJK}$. We argue that the action should contain terms with odd number of $M^{IJL}$ as well. We modify  the action to include them. 
\vfill \setcounter{page}{0} \setcounter{footnote}{0}
\newpage

%\section{Introduction} \label{intro}
Following the idea that the Chern-Simons gauge theory may be used to describe the dynamics of coincident M2-branes \cite{Schwarz:2004yj}, Bagger and Lambert \cite{Bagger:2007vi} as well as Gustavsson \cite{Gustavsson:2007vu} have constructed three dimensional ${\mathcal{N}}=8$ superconformal $SO(4)$ Chern-Simons gauge theory based on 3-algebra.
It is believed that the BLG world volume theory at level one describes two M2-branes on $R^8/Z_2$ orbifold \cite{Lambert:2008et}. The world volume theory of N M2-branes on $R^8/Z_k$ orbifold has been constructed in \cite{Aharony:2008ug} which is given by ${\mathcal{N}}=6$ $U(N)_k\times U(N)_{-k}$ Chern-Simons gauge theory.

The signature of the metric on the 3-algebra in the BLG model is positive definite. This assumption has been relaxed in \cite{Gomis:2008uv} to study $N$ coincident M2-branes  in flat spacetime. The so called BF membrane theory with arbitrary semi-simple Lie group has been proposed in \cite{Gomis:2008uv}. This theory has ghost fields, however, there are different arguments that model may be unitary due to the particular form of the interactions \cite{Gomis:2008uv,Cecotti:2008qs}.  The bosonic part of the Lagrangian for gauge group $U(N)$ is given by\footnote{Our index convention is that
 $a,b,...=0,1,2$;  
$i,j,...=3,...,9$ and $I,J,...=3,4,..., 10$.}
\beqa
L=\Tr\left(\frac{1}{2}\epsilon^{abc}B_aF_{bc}-\frac{1}{2}\hD_aX^I\hD^aX^I+\frac{1}{12}M^{IJK}M^{IJK}\right)+(\prt_aX^I_--\Tr(B_aX^I))\prt^aX^I_+\labell{L1}
\eeqa
where $A_a,B_a,X^I$ are in adjoint representation of $U(N)$ and $X^I_-,X^I_+$ are singlet under $U(N)$, and
\beqa
M^{IJK}&\equiv&X^I_+[X^J,X^K]+X^J_+[X^K,X^I]+X^K_+[X^I,X^J]\labell{MIJK}\\
\hD_aX^I&=&D_aX^I-X^I_+B_a\,,\qquad D_aX^I=\prt_aX^I+i[A_a,X^I]\nonumber
\eeqa
Obviously the above Lagrangian  is invariant under global $SO(8)$ transformation  and under $U(N)$ gauge transformation associated with the $A_a$ gauge field. It is also invariant under gauge transformation associated with the $B_a$ gauge field
\beqa
\delta_B X^I=X^I_+\Lam\,,\qquad \delta_B B_a=D_a\Lam\,,\qquad \delta _BX^I_+=0\,,\qquad \delta _BX^I_-=\Tr(X^I\Lam)\labell{Bt}
\eeqa
The Lagrangian \reef{L1} is a candidate to describe the dynamics of N M2-branes in flat supergravity background. 
%A nonlinear extension of this Lagrangian is proposed in \cite{Iengo:2008cq}.

The equation of motion for $X^I_-$ gives $\prt_a\prt^aX^I_+=0$. Following the procedure found in \cite{Mukhi:2008ux}, if one of the scalars $X^I_+$ takes large expectation value, \ie  $X^I_+=g_{YM}\delta^{I10}$, and  the gauge symmetry \reef{Bt} is fixed by setting $X^{10}=0$, then the above action reduces to 
\beqa
L&=&\Tr\left(\frac{1}{2}\epsilon^{abc}B_aF_{bc}-\frac{1}{2}D_aX^iD^aX^i-\frac{g_{YM}^2}{2}B_aB^a-\frac{g_{YM}^2}{4}[X^i,X^j][X^j,X^i]\right)
\eeqa
Under the dNS duality transformation \cite{Nicolai:2003bp}
\beqa
\Tr\left(\frac{1}{2}\epsilon^{abc}B_aF_{bc}-\frac{g_{YM}^2}{2}B_aB^a\right)&\rightarrow &\Tr\left(-\frac{1}{4g_{YM}^2}F^{ab}F_{ab}\right)
\eeqa
the action becomes identical to the low energy action of N D2-branes in flat spacetime. The nonlinear extension of this action is given by the  following nonabelian  DBI action \cite{Tseytlin:1997csa,Myers:1999ps}: 

 \beqa
S^{D_2}&=&-T_2\int
d^{3}\sigma \STr\left(\sqrt{\det(Q)}\right.\labell{nonab}\\
&&\times\left.
\sqrt{-\det\left(\frac{}{}\eta_{ab}+\la^2 g_{YM}^2D_aX^i(Q^{-1})^{ij}D_bX^j
+\la F_{ab}\right)} \right)\,\,,\nonumber \eeqa 
where the matrix $Q^{ij}$ is  \beqa
Q^{ij}&=&I\delta^{ij}+i\la g_{YM}^2[X^i,X^j] \nonumber
\eeqa where $\la\equiv 2\pi\alpha'$. Here  the transverse scalars in \cite{Myers:1999ps} are normalized as  $\Phi^i=g_{YM}\la X^i$. The trace in the action 
is completely symmetric between all  matrices
 $F_{ab},DX^i, [X^i,X^j]$. The D2-brane tension is $
T_{2}=1/(2\pi)^2\ell_s^3g_s$.
The 3-dimensional Yang-Mills coupling constant is related to the tension of D2-brane as $\la^2T_2=1/g_{YM}^2$. 

An extension of the above action to the action of N M2-branes has been proposed in \cite{Iengo:2008cq}\footnote{ See \cite{Kluson:2008nw}  for nonlinear action of M2-brane in abelian case.}. This action is given as
\beqa
&\!\!\!\!\!&-T_2\STr\left(\sqrt{-\det\left(\eta_{ab}+\frac{1}{T_2}\tD_aX^I(\tQ^{-1})^{(IJ)}\tD_bX^J\right)}(\det(\tQ))^{1/4}\right)+\Tr\left(\frac{1}{2}\epsilon^{abc}B_aF_{bc}\right)\labell{LN}\\
&&+(\prt_aX^I_--\Tr(B_aX^I))\prt^aX^I_+-\Tr\left(\frac{X_+\inn X}{X^2_+}\hD_aX^I\prt^aX^I_+-\frac{1}{2}\left(\frac{X_+\inn X}{X^2_+}\right)^2\prt_aX^I_+\prt^aX^I_+\right)\nonumber
\eeqa
where $X_+^2=X^I_+X^I_+$ and
\beqa
\tD_aX^I&=&\hD_aX^I-\left(\frac{X_+\inn X}{X^2_+}\right)\prt_aX^I_+\labell{tDX}\\
\tQ^{IJ}&=&s^{IJ}+\left(\frac{X_+^I X^J_+}{X^2_+}\right)(\det(s)-1)\,,\qquad s^{IJ}=\delta^{IJ}+\frac{i}{\sqrt{T_2}}\frac{X_+^KM^{IJK}}{\sqrt{X^2_+}}\nonumber
\eeqa
The M2-brane tension is $T_{2}=1/(2\pi)^2\ell_p^3$. It is shown in \cite{Iengo:2008cq} that this BF type nonlinear Lagrangian reduces to \reef{L1} at low energy. Since only the symmetric part of matrix $\tQ^{-1}$ appears in the action, it includes only terms with even number of $M^{IJK}$.

 In this paper we would like to modify this  action to includes terms with even and odd number of $M^{IJK}$. To this end, we back to the original action \reef{nonab} from which the above action has been found. 
The matrix $(Q^{-1})^{ij}$ in the D2-branes action has the following expansion:
\beqa
(Q^{-1})^{ij}&=&\delta^{ij}-i\la g_{YM}^2[X^i,X^j]+(i\la g_{YM}^2)^2[X^i,X^k][X^k,X^j]\nonumber\\
&&-(i\la g_{YM}^2)^3[X^i,X^k][X^k,X^l][X^l,X^j]+\cdots
\eeqa
Using the fact that all brackets appear in symmetric form in the action, one can easily check that all terms with even number of brackets are symmetric in $i,j$ and all terms with odd number of brackets are antisymmetric. It can be written as  $(Q^{-1})^{ij}=R^{ij}+\la P^{ij}$ where the symmetric matrix $R_{ij}$ and antisymmetric matrix $P_{ij}$ are
\beqa
R^{ij}&=&(q^{-1})^{ij}\,,\qquad
P^{ij}\,\,=\,\,\left(-i g_{YM}^2[X^i,X^n]\right)(q^{-1})^{nj}\nonumber\\
q^{ij}&=&\delta^{ij}+\la^2g_{YM}^4[X^i,X^k][X^k,X^j]
\eeqa
Note that $1/\det(R)=\det(q)=(\det(Q))^2$. 

Both $R^{ij}$ and $P^{ij}$  contributes to  action \reef{nonab}. For example, if one considers the $\delta^{ij}$ term of $R_{ij}$, then it is obvious that its contribution is not zero. To see that the antisymmetric matrix $P^{ij}$ has contribution, consider for instance $[X^i,X^j]$ term of $P^{ij}$. The expansion of the square root in  action \reef{nonab} has among other things  the following term:
\beqa
\STr(D_aX^i[X^i,X^j]D_bX^jF_{ba})=\frac{1}{2}\left(\Tr(D_aX^i[X^i,X^j]D_bX^j)+\Tr(D_aX^iD_bX^j[X^i,X^j])\right)F_{ba}\nonumber
\eeqa
where we have assumed on the right hand side that $F_{ab}$ is abelian.  Using the fact that $F_{ab}=-F_{ba}$, one observes that both terms on the right hand side are equal. Hence, $P_{ij}$ has non-vanishing contribution to the action. 
The symmetric trace makes the matrix $\eta_{ab}+\la^2 g_{YM}^2D_aX^iR^{ij}D_bX^j$ in the action to be symmetric and matrix $\la^2 g_{YM}^2D_aX^iP^{ij}D_bX^j$ to be antisymmetric. 

Now we use the following duality transformation \cite{Iengo:2008cq}:
\beqa
-T_2\sqrt{-\phi\det(g_{ab}+\la F_{ab})}&\rightarrow &-T_2\sqrt{-\phi\det(g_{ab}+\frac{g_{YM}^2}{T_2}\frac{B_aB_b}{\phi})}+\frac{1}{2}\epsilon^{abc}B_aF_{bc}\labell{iden}
\eeqa
for any scalar $\phi$, any symmetric matrix $g_{ab}$ and any antisymmetric matrix $F_{ab}$.
The symmetric trace prescription   allows us to use  the  matrix extension of the above identity  in which the $\STr$ appears in both sides.  

Using the above duality transformation,  the  action \reef{nonab} can be written in the following form:
 \beqa
S^{D_2}&=&-T_2\int
d^{3}\sigma \STr\left(\sqrt{\det(Q)}\right.\labell{nonab1}\\
&&\times\left.
\sqrt{-\det\left(\eta_{ab}+\frac{1}{T_2}D_aX^iR^{ij}D_bX^j+\frac{g_{YM}^2}{T_2}\frac{B_aB_b}{\det(Q)}
\right)} \right)\nonumber\\
&&+\frac{1}{2}\int
d^{3}\sigma \STr\left(\frac{}{}\epsilon^{abc}B_a(F_{bc}+\frac{1}{T_2}D_bX^iP^{ij}D_cX^j)\right)\,\,,\nonumber \eeqa 
where $B_a$ is a matrix in the adjoint representation of $U(N)$.

%\section{Multiple M2-branes effective  action}

Using the prescription given in \cite{Mukhi:2008ux}, one expects that  M2-branes effective   action should be reduced to the  D2-branes   action when $X^I_+$  takes a  large expectation value. 
Following \cite{Iengo:2008cq}, the M2-branes  extension of  \reef{nonab1} should have $SO(8)$ invariant term $\tD_aX^{I}\tR^{IJ}\tD_bX^{J}$ where $\tD_aX^{I}$ and $\tR_{IJ}$ should be defined to be invariant under the $B_a$ gauge transformation \reef{Bt} and   when $X^I_+=v\delta^{I10}$ where $v=g_{YM}$, they should satisfy the boundary condition \cite{Iengo:2008cq}: 
\beqa
 \tD_aX^{I}\tR^{IJ}\tD_bX^{J}\rightarrow D_aX^{i}R^{ij}D_bX^{j}+v^2\frac{B_aB_b}{\det(Q)}
 \eeqa
This makes $\tD_aX^I$ to be defined as in  \reef{tDX}, and 
the boundary value of $\tR_{IJ}$ to be \cite{Iengo:2008cq}
\beqa
\tR^{ij}=R^{ij}\qquad,\qquad \tR^{i10}=\tR^{10i}=0\qquad,\qquad \tR^{1010}=\frac{1}{\det(Q)}\labell{boundary}
\eeqa

 An ansatz  for $\tR^{IJ}$ which is consistent with the above boundary condition may be 
 \beqa
 \tR^{IJ}=(S^{-1})^{IJ}+aX^I_+X^J_+
 \eeqa
 where $a$ is a $SO(8)$ invariant term which can  be found from the above boundary condition, and
 \beqa
 S^{IJ}&=&\delta^{IJ}-\frac{1}{T_2}M^{IKM}M^{JKN}\left(\frac{X^M_+X^N_+}{X^2_+}\right)
 \eeqa
 Assuming that $M^{IJK}$ should appear in symmetric form in the M2-branes action which is inherited from D2-branes action in which  $[X^i,X^j]$ appears in symmetric form, one observes that matrix $S^{IJ}$ is symmetric. This matrix satisfies the boundary $S^{ij}=q^{ij}$, $S^{i10}=S^{10i}=0$ and $S^{1010}=1$. So at the boundary $\det(S)=\det(q)=(\det(Q))^2$.
Imposing the boundary condition, one finds
\beqa
\tR^{IJ}&=&(S^{-1})^{IJ}+\frac{X^I_+X^J_+}{X^2_+}\left(\frac{1}{\sqrt{\det(S)}}-1\right) \nonumber\eeqa
This symmetric matrix has even number of $M^{IJK}$. Note also $\tR^{IJ}=\delta^{IJ}+O(\frac{1}{T_2})$.

%Another ansatz  for $\tR^{IJ}$  may be 
% \beqa
% \tR^{IJ}=(S^{-1})^{IJ}
% \eeqa
% where 
% \beqa
% S^{IJ}&=&\delta^{IJ}+aM^{IKM}M^{JKM}
% \eeqa
%Imposing the boundary condition $\tR^{ij}=q^{ij}$, one finds $a=-1/(2T_2)$. 

The M2-branes action should also have $SO(8)$ invariant term $\tD_aX^{I}\tP^{IJ}\tD_bX^{J}$ where   $\tP^{IJ}$ should be defined to  satisfy  the boundary condition: 
\beqa
 \tD_aX^{I}\tP^{IJ}\tD_bX^{J}\rightarrow D_aX^{i}P^{ij}D_bX^{j}
 \eeqa
 This fixes $\tP^{IJ}$ to be
 \beqa
 \tP^{IJ}&=&iM^{IKN}(S^{-1})^{NJ}X^K_+
 \eeqa
 This matrix has odd number of $M^{IJL}$. The symmetric trace prescription then makes it to be antisymmetric matrix.
 The first term in the third  line of \reef{nonab1} is invariant under the gauge transformations associated with $B_a$ and $A_a$ fields. The second term  however is not invariant under these  gauge transformations. With the above modification, the second  term  is not yet invariant under the gauge transformation \reef{Bt}. To make it invariant,  we have to also extend $B_aX_+^K$ to $-\tD_aX^K$
 %\beqa
 %B_a+\frac{\prt_aX_+\inn X-X_+\inn D_aX}{X^2_+}
 %\eeqa
 which is invariant under the gauge transformation \reef{Bt}.
 % and  reduces to $B_aX_+^K$ at the boundary after fixing the $B_a$ gauge symmetry by setting $X^{10}=0$.
 
 So, a proposal for the extension of action $S^{D_2}$ to M2-branes  may be  given by the following Lagrangian:
\beqa
L_1^{M_2}&=&-T_2\STr\left(\left(\det(S)\right)^{1/4}
\sqrt{-\det\left(\eta_{ab}+\frac{1}{T_2}\tD_aX^I\tR^{IJ}\tD_bX^J
\right)} \right)\labell{LM}\\
&&+\frac{1}{2}\epsilon^{abc}\left(\Tr(B_aF_{bc})-\frac{i}{T_2}\STr\left(\tD_aX^K\tD_bX^{I}M^{IKN}(S^{-1})^{NJ}\tD_cX^{J}\right)\right)\nonumber\\
&&+(\prt_aX^I_--\Tr(B_aX^I))\prt^aX^I_+-\Tr\left(\frac{X_+\inn X}{X^2_+}\hD_aX^I\prt^aX^I_+-\frac{1}{2}\left(\frac{X_+\inn X}{X^2_+}\right)^2\prt_aX^I_+\prt^aX^I_+\right)\nonumber \eeqa 
 The  symmetric trace in the first two lines is  between gauge invariants $\tD_aX^I$ and $M^{IJK}$. This action is manifestly invariant under global $SO(8)$ transformation and is invariant under gauge transformation associated  with gauge fields $A_a$ and $B_a$. As has been discussed in \cite{Iengo:2008cq}, the last line should be  added to have consistency with the low energy action.
 
 The equation of motion for $X^I_-$ gives $\prt_a\prt^a X^I_+=0$. As in \cite{Iengo:2008cq}, if one of the scalars $X^I_+$ takes large expectation value, \ie $X^I_+=v\delta^{I10}$, then $\tD_aX^{i}=D_aX^{i}$ and $\tD_aX^{10}=D_aX^{10}-vB_a$. Now fixing  the gauge symmetry \reef{Bt}  by setting $X^{10}=0$, one recovers the D2-branes action \reef{nonab1}. On the other hand, if the shift symmetry  $X^I_-\rightarrow X^I_-+c^I$ is gauged as in \cite{Bandres:2008kj,Ezhuthachan:2008ch} by introducing a new field $C_a^I$ and writing $\prt_aX^I_-$ as $\prt_aX^I_--C_a^I$, then equation of motion for the new field  gives $\prt_a X^I_+=0$ which has only constant solution $X^{I}_+=v^I$. Using the $SO(8)$ symmetry, one can write it as $X^{I}_+=v\delta^{I10}$. Then the above theory would be classically equivalent to the $D_2\bar{D}_2$ theory \reef{nonab1}. 
 
 Using the identity \cite{Iengo:2008cq}
 \beqa
 M^{IKM}M^{IKN}\left(\frac{X^M_+X^N_+}{X^2_+}\right)&=&\frac{1}{3}M^{IJK}M^{IJK}
 \eeqa
 one finds
 \beqa
 \left(\det(S)\right)^{1/4}&=&1-\frac{1}{12T_2}M^{IJK}M^{IJK}+\cdots
 \eeqa
 So at low energy the action \reef{LM}  reduces to \reef{L1}. The action \reef{LM} has terms with even and odd number of $M^{IJK}$. The terms with even number of $M^{IJK}$ are those appear in \reef{LN}. However, the terms with odd number of $M^{IJK}$ which are non-vanishing, \ie 
 \beqa
 \frac{i}{2T_2}\epsilon^{abc}\STr\left(\tD_aX^K\tD_bX^{I}M^{IKN}(S^{-1})^{NL}\tD_cX^{J}\right)
 \eeqa
are new couplings that are not included in \reef{LN}.

 Another  ansatz  for $\tR^{IJ}$  which may be consistent with the boundary condition \reef{boundary} is 
 \beqa
 \tR^{IJ}=(s^{-1})^{IJ}
 \eeqa
where 
 \beqa
 s^{IJ}&=&\delta^{IJ}+aM^{IKM}M^{JKM}
 \eeqa
in which $a$ is a constant. Imposing the boundary condition $\tR^{ij}=R^{ij}$, one finds $a=-1/(2T_2)$. Interestingly, it also satisfies the boundary condition $(s^{-1})^{1010}=1/\det(Q)$. To see this, consider the expansion of $(s^{-1})^{IJ}$
\beqa
(s^{-1})^{IJ}&=&\delta^{IJ}-a(MM)^{IJ}+a^2(MMMM)^{IJ}-a^3(MMMMMM)^{IJ}+\cdots
\eeqa
where our notation is such that $(MMMM)^{IJ}=M^{IKM}M^{NKM}M^{NPQ}M^{JPQ}$. Using the definition of $M^{IJK}$ in \reef{MIJK}, one finds that the $1010$ component of this matrix is
\beqa
(s^{-1})^{1010}&=&\frac{1}{1+aM^{10ij}M^{10ij}}=\frac{1}{1+\frac{g_{MY}^2}{T_2}[X^i,X^j][X^j,X^i]}
\eeqa
On the other hand, because of  the symmetric trace in the action, one finds $\det(Q)=1+\frac{g_{MY}^2}{T_2}[X^i,X^j][X^j,X^i]$. 
At the boundary, one also has the relation $\det(s)=(\det(Q))^3$.

So, another  proposal for the extension of action $S^{D_2}$ to M2-branes  may be  given by the following Lagrangian:
\beqa
L_2^{M_2}&=&-T_2\STr\left(\left(\det(s)\right)^{1/6}
\sqrt{-\det\left(\eta_{ab}+\frac{1}{T_2}\tD_aX^I(s^{-1})^{IJ}\tD_bX^J
\right)} \right)\labell{LM2}\\
&&+\frac{1}{2}\epsilon^{abc}\left(\Tr(B_aF_{bc})-\frac{i}{T_2}\STr\left(\tD_aX^K\tD_bX^{I}M^{IKN}(s^{-1})^{NJ}\tD_cX^{J}\right)\right)\nonumber\\
&&+(\prt_aX^I_--\Tr(B_aX^I))\prt^aX^I_+-\Tr\left(\frac{X_+\inn X}{X^2_+}\hD_aX^I\prt^aX^I_+-\frac{1}{2}\left(\frac{X_+\inn X}{X^2_+}\right)^2\prt_aX^I_+\prt^aX^I_+\right)\nonumber \eeqa 
In this case also, using the expansion
\beqa
 \left(\det(s)\right)^{1/6}&=&1-\frac{1}{12T_2}M^{IJK}M^{IJK}+\cdots
 \eeqa
 one observes that  at low energy the above action   reduces to \reef{L1} as expected. For constant $X^I_+$, the above action produces  the couplings found in \cite{Alishahiha:2008rs} at order $1/T_2$.
% {\bf Acknowledgement}: I would like to thank K. Hashimoto for discussion.

%\end{document}
%\newpage

\end{document}